\documentclass[preprint,aps,floatfix,showpacs,nofootinbib]{revtex4}
\usepackage{dcolumn}
\usepackage{bm}
\usepackage{color}
\usepackage{graphicx}
\usepackage{epsf}
\usepackage{pstricks}
\usepackage{slashed}
\usepackage{amssymb}
\usepackage{array}
\usepackage{multirow}

\begin{document}

\title{Local chiral potentials and the structure of light nuclei}
\author{M.\ Piarulli$^{\rm a}$, L.\ Girlanda$^{\rm b,c}$, R.\ Schiavilla$^{\,{\rm d,e}}$, A.\ Kievsky$^{\rm f}$, 
A.\ Lovato$^{\rm a}$, L.E.\ Marcucci$^{\rm g,f}$, Steven C.\ Pieper$^{\rm a}$,\\
 M.\ Viviani$^{\rm f}$, and R.B.\ Wiringa$^{\rm a}$}
\affiliation{
$^{\,{\rm a}}$\mbox{Physics Division, Argonne National Laboratory, Argonne, Illinois 60439, USA}	\\
$^{\,{\rm b}}$\mbox{Department of Mathematics and Physics, University of Salento, 73100 Lecce, Italy}\\
$^{\,{\rm c}}$\mbox{INFN-Lecce, 73100 Lecce, Italy}\\
$^{\,{\rm d}}$\mbox{Theory Center, Jefferson Lab, Newport News, VA 23606, USA}\\
$^{\,{\rm e}}$\mbox{Department of Physics, Old Dominion University, Norfolk, VA 23529, USA}\\
$^{\,{\rm f}}$\mbox{INFN-Pisa, 56127 Pisa, Italy}\\
$^{\,{\rm g}}$\mbox{Department of Physics, University of Pisa, 56127 Pisa, Italy}
}
\date{\today}

\begin{abstract}
We present fully local versions of the minimally non-local nucleon-nucleon potentials
constructed in a previous paper [M.\ Piarulli {\it et al.}, Phys.\ Rev.\ C {\bf 91}, 024003
(2015)], and use them in hypersperical-harmonics and quantum Monte Carlo calculations
of ground and excited states of $^3$H, $^3$He, $^4$He, $^6$He, and $^6$Li nuclei.
The long-range part of these local potentials includes one- and two-pion exchange
contributions without and with $\Delta$-isobars in the intermediate states up to order
$Q^3$ ($Q$ denotes generically the low momentum scale) in the chiral expansion, 
while the short-range part consists of contact interactions up to
order $Q^4$.  The low-energy constants multiplying these contact interactions are fitted 
to the 2013 Granada database in two different ranges of laboratory energies, either 0--125
MeV or 0--200 MeV, and to the deuteron binding energy and $nn$ singlet scattering length. 
Fits to these data are performed for three models characterized by long- and
short-range cutoffs, $R_{\rm L}$ and $R_{\rm S}$ respectively, ranging from
$(R_{\rm L},R_{\rm S})=(1.2,0.8)$ fm down to $(0.8,0.6)$ fm.  The long-range (short-range)
cutoff regularizes the one- and two-pion exchange (contact) part of the potential.
\end{abstract} 


\index{}\maketitle

\section{Introduction}
\label{sec:intro}
The understanding of the structure and reactions 
of nuclei and nuclear matter has been a 
long-standing goal of nuclear physics.  In this respect,
few- and many-body systems provide
a laboratory for studying nuclear forces with a
variety of numerical and computational techniques. 
In recent years, rapid advances in {\it ab initio} few- and
many-body methods, such as no-core shell model (NCSM)~\cite{Barrett13,Jurgenson13}, 
coupled cluster (CC)~\cite{Hagen14,Hagen14_1} and
hyperspherical harmonics (HH)~\cite{Viviani06,Marcucci09,Viviani09,Viviani10} expansions,
similarity renormalization group (SRG) approaches~\cite{Bogner10,Hergert13},
self-consistent Green's function techniques~\cite{Dickhoff04,Soma13}, and
quantum Monte Carlo (QMC) methods~\cite{Carlson14}, in combination with the rapid increase
in computational resources, have made it possible to test
conventional theories and new ones, such as chiral effective field theory ($\chi$EFT),
in calculations of nuclear structure and reactions.

During the last quarter century, $\chi$EFT,
originally proposed by Weinberg in the early 1990's~\cite{Weinberg90},
has been widely used for the derivation of nuclear forces and electroweak currents.
Such a theory provides the most general scheme accommodating 
all possible interactions among nucleons, $\Delta$ isobars, and pions compatible
with the relevant symmetries---in particular chiral symmetry---of low-energy
quantum chromodynamics (QCD), the underlying theory of strong
interactions. By its own nature, $\chi$EFT is organized within a given
power counting scheme and the resulting chiral potentials (and currents)
are systematically expanded in powers of $Q/\Lambda_\chi$ with 
$Q\ll\Lambda_\chi$, where $Q$ denotes generically a low momentum
and $\Lambda_\chi \sim 1$  GeV specifies the chiral-symmetry breaking scale
(see Refs.~\cite{Epelbaum09r,Entem11} for recent review articles).

The power counting of $\chi$EFT indicates that nuclear forces are dominated by
nucleon-nucleon ($NN$) interactions, a feature which was already known 
before $\chi$EFT was introduced but could be justified more formally 
with the advent of such a theory~\cite{Weinberg90}. Many-body forces are
suppressed by powers of $Q$; however, 
the inclusion of three-nucleon forces ($3N$) is mandatory at the level of accuracy
now reached by few- and many-body calculations (see~\cite{Kalantar12,Hammer13} and
references therein for a comprehensive review on this topic).
Being the dominant contribution of the nuclear forces, a great 
deal of attention has been devoted to the derivation and optimization of 
$NN$ interactions. 

About a decade ago, $NN$ interactions up to 
next-to-next-to-next-to-leading order (N3LO or $Q^4$) in the chiral expansion were 
derived~\cite{Kaiser97,Kaiser98,Kaiser99,Kaiser00,
Kaiser01,Kaiser011,Kaiser02,Epelbaum9800,Entem02,Krebs07} 
and quantitative $NN$ potentials were developed~\cite{Entem03,Epelbaum05}
at that order. These N3LO $NN$ interactions are separated into 
pion-exchange contributions and contact terms.
Pion-exchange contributions represent the long-range part of the $NN$ interactions
and include at leading order (LO or $Q^0$)
the well-known static one-pion-exchange (OPE) potential and
at higher orders, namely next-to-leading (NLO or $Q^2$), 
next-to-next-to-leading (N2LO or $Q^3$) and N3LO, the two-pion-exchange
(TPE) potential due to leading and sub-leading
$\pi N$ couplings. These sub-leading chiral constants can consistently be obtained from low-energy 
$\pi N$ scattering data~\cite{Krebs07,Siemens16,Yao16}. 
Also three-pion-exchange ($3\pi$) shows up for the
first time at N3LO; in Refs.~\cite{Kaiser99,Kaiser00}, it was demonstrated
that the $3\pi$ contributions at this order are negligible.
More recently two- and three-pion
exchange contributions that occur at N4LO ($Q^5$)~\cite{Entem15,Epelbaum15} and 
N5LO ($Q^6$)~\cite{Entem2015} have been investigated.

Contact terms encode the short-range physics, and their strength 
is specified by unknown low-energy constants (LECs).
In order to fix these LECs, $NN$ chiral potentials have been confronted with the $pp$ and $np$ scattering databases
up to lab energy of 300 MeV.  These databases have been provided by the Nijmegen group~\cite{Stoks93,Stoks94},
the VPI/GWU group~\cite{SM99}, and more recently the Granada group~\cite{Navarro1314}.
In the standard optimization procedure the potentials are first constrained 
by fitting $np$ and $pp$ phase shifts, and then the fit is refined by minimizing the total $\chi^2$ obtained
from a direct comparison with the $NN$ scattering data.
Entem and Machleidt~\cite{Entem03} used their N3LO chiral
potential to fit $pp$ and $np$ scattering data in the Nijmegen database
up to laboratory energy of 290 MeV with a total $\chi^2$/datum of 1.28.
Other available chiral potentials~\cite{Epelbaum05} have not
been fitted to scattering data directly but rather to phase shifts obtained
in the Nijmegen analysis (the recent upgrade~\cite{Epelbaum15}
of Ref.~\cite{Epelbaum05} relies on this procedure, while
in Refs.~\cite{Entem15,Entem2015} a study of peripheral phase shifts is carried out
with two- and three-pion exchange contributions up to order $Q^5$ and $Q^6$, respectively).

Recently, a different optimization strategy has been introduced 
by A. Ekstrom {\it et al.}~\cite{Ekstrom15}. This new approach is based on
a simultaneous fit of the $NN$ and $3N$ forces to low-energy
$NN$ data, deuteron binding energy, and
binding energies and charge radii of hydrogen, helium,
carbon, and oxygen isotopes. These authors considered
the $NN+3N$ interaction at N2LO, namely N2LO$_{\rm sat}$, where the $NN$ sector is
constrained by $pp$ and $np$ scattering observables from the SM99 database up to 35 MeV
scattering energy in the laboratory system with a total $\chi^2$/datum $\approx$ 4.3.\\

The family of  $NN$ chiral interactions mentioned above
are formulated in momentum-space and have the feature 
of being strongly non-local in coordinate space, making them not well-suited for certain
numerical algorithms, for example QMC.  Up to until recently, QMC methods, 
such as variational Monte Carlo (VMC), Green's function Monte Carlo (GFMC) and auxiliary 
field diffusion Monte Carlo (AFDMC), have been used to compute 
the properties of light nuclei with mass number $A \leq 12$, closed shell
nuclei $^{16}$O and $^{40}$Ca, and nucleon matter by using phenomenological nuclear Hamiltonians based
on the Argonne $v_{18}$ (AV18) two-nucleon potential~\cite{Wiringa95}
and the Urbana/Illinois (U/IL) series of three-nucleon potentials~\cite{Carlson83,Pudliner96,Pieper08,Pieper01}. 
While QMC has had great success in predicting many
nuclear properties, such as spectra, electromagnetic form
factors, electroweak transitions, low-energy scattering and response,
nevertheless it has been limited to realistic Hamiltonians based on the
AV18 and U/IL models and other simpler local interactions. The reason is that 
local coordinate-space interactions are particularly convenient for 
QMC techniques, and the AV18 and U/IL models fall into
this category, while many of the available $NN$ chiral interactions have
strong non-localities. These non-localities come about because
of (i) the specific choice made to regularize the momentum space potential,
and (ii) contact interactions that depend not only on the 
momentum transfer ${\bf k}={\bf p}^\prime -{\bf p}$ but also 
on ${\bf K}=({\bf p}^\prime +{\bf p})/2$ (${\bf p}$ and ${\bf p}^\prime$ 
are the initial and final relative momenta of the two nucleons).

Local chiral interactions were developed
up to N2LO (or $Q^3$)~\cite{Gezerlis13,Gezerlis14} only recently.
These interactions are regularized in coordinate space by a cutoff
depending only on the relative distance between the two nucleons,
and use Fierz identities to remove completely the dependence on the
relative momentum $-i\, {\bm \nabla}$ (or equivalently ${\bf K}$),
by selecting appropriate combinations of contact operators.
The LECs multiplying these contact terms have
been fixed by performing $\chi^2$ fits to the $np$ phase shifts from
the Nijmegen partial-wave analysis (PWA) up to 150 MeV lab energy.
The resulting chiral potentials have been used in GFMC calculations
for $A \leq 5$ nuclei and AFDMC calculations of neutron 
matter~\cite{Gezerlis14,Lynn14,Lynn16}.
While this Fierz re-arrangement is effective in completely removing 
non-localities at N2LO, it cannot do so at N3LO.
As shown in Ref.~\cite{Piarulli15}, operator structures depending quadratically
on $-i\, {\bm \nabla}$ are unavoidable, and therefore
the potentials constructed in Ref.~\cite{Piarulli15} belong to the
class of ``minimally non-local'' chiral potentials at N3LO.

In the present work we construct fully local versions of these minimally 
non-local $NN$ potentials~\cite{Piarulli15} by dropping the terms proportional
to ${\bm \nabla}^2$, and use them in HH, VMC
and GFMC calculations of ground and excited states of $^3$H, $^3$He,
$^4$He, $^6$He, and $^6$Li nuclei.  The paper is organized as follows. 
In the next section we summarize the main points of Ref.~\cite{Piarulli15},
and then proceed to discuss the modifications adopted in this work in order
to construct the new class of local potentials.  In Sec.~\ref{sec:phases} we
provide the $\chi^2$ values obtained by performing different types of fits, show
the calculated phase shifts for the lower partial waves (S, P, and D waves), 
and compare these phase shifts to those from recent PWA's.
There we also provide tables of the $pp$, $np$
and $nn$ effective range parameters and deuteron properties.
In Sec.~\ref{sec:nuclei} the HH, VMC and GFMC methods are briefly described
and results for the binding energies of $A\,$=$\, 3$, 4, and 6 nuclei are discussed.
Clearly, the N3LO calculations reported here with only two-body forces
are incomplete, since three-body forces start to come in at N2LO.  Nevertheless,
they provide the basis for the calculations of light nuclei structure based
on chiral two- {\it and} three-body forces which will follow.
\section{Local Chiral $NN$ Potentials}
\label{sec:pots}
Following Ref.~\cite{Piarulli15}, the local $NN$ potential constructed
in the present work is written as 
a sum of an electromagnetic-interaction
component, $v_{12}^{\rm EM}$,
and a strong-interaction component, $v_{12}$.
The $v_{12}^{\rm EM}$ interaction is 
the same as that used in the AV18 potential~\cite{Wiringa95}, while
the $v_{12}$ one is obtained in $\chi$EFT
and is conveniently separated 
into long- and short-range 
parts, respectively $v_{12}^{\rm L}$ and $v_{12}^{\rm S}$. 
The $v_{12}^{\rm L}$ part includes the 
one-pion-exchange (OPE) and 
two-pion-exchange (TPE) contributions up to 
N2LO (or $Q^3$) in the
chiral expansion. The TPE component also contains
diagrams involving $\Delta$-isobars in
intermediate states~\cite{Piarulli15}.

The strength of this long-range part is fully 
determined by the nucleon and nucleon-to-$\Delta$ axial
coupling constants $g_A$ and $h_A$, 
the pion decay amplitude $F_\pi$, and the
sub-leading N2LO LECs $c_1$, $c_2$, $c_3$, $c_4$, 
and $b_3+b_8$, constrained by
reproducing $\pi N$ scattering data~\cite{Krebs07}.
Note that the LEC $(b_3 + b_8)$ is explicitly retained 
in our fitting procedure, even though it has been shown to be 
redundant at this order~\cite{Long11}.
Here and in what follows, we adopt the same values for pion
and nucleon masses, $F_\pi$, $g_A$ and $h_A$ and
the sub-leading N2LO LECs as listed in Tables I and II
of Ref.~\cite{Piarulli15}.

The potential $v_{12}^{\rm L}$ 
can be written in coordinate space 
as a sum of 8 operators,
\begin{equation}
\label{eq:vlr}
v_{12}^{\rm L}= \left[\sum_{l=1}^{6} v_{\rm L}^l (r) \, O^l_{12}\right]
+v_{\rm L}^{\sigma T}(r)  \, O^{\sigma T}_{12} + v_{\rm L}^{t T}(r)  \, O^{t T}_{12} \ ,
\end{equation}
where 
\begin{equation}
O^{l=1,...,6}_{12}=\left[{\bf 1}\, ,\, {\bm \sigma}_1\cdot {\bm \sigma}_2\, , \,S_{12}\right]
\otimes\left[{\bf 1}\, ,\, {\bm \tau}_1\cdot {\bm \tau}_2\right] \ ,
\end{equation}
$O^{\sigma T}_{12}={\bm \sigma}_1\cdot {\bm \sigma}_2\,T_{12}$, and
$O^{tT}_{12}=S_{12}\,T_{12}$, and $T_{12}=3\,\tau_{1z}\tau_{2z}-{\bm
  \tau}_1\cdot {\bm \tau}_2$ is the isotensor operator. 
The first 6 terms (the so-called $v_6$ operator structure) 
in Eq.~(\ref{eq:vlr}) are the charge-independent 
(CI) central, spin, and tensor components without and with
the isospin dependence ${\bm \tau}_1\cdot {\bm \tau}_2$,
while the last two terms (proportional to $T_{12}$) are
the charge-independence breaking (CIB) central and tensor
components induced by the difference between the neutral 
and charged pion masses in the OPE.
The radial functions $v_{\rm L}^l(r)$, 
$v_{\rm L}^{\sigma T}(r)$, and $v_{\rm L}^{t T}(r)$ are 
explicitly given in Appendix A of~\cite{Piarulli15}. 
The singularities at the origin are regularized 
by cutoff functions of the form
\begin{equation}
\label{eq:ctff}
 C_{R_{\rm L}}(r)=1-\frac{1}{(r/R_{\rm L})^6 \,  e^{(r-R_{\rm L})/a_{\rm L}} +1} \ , 
\end{equation}
where three values for the radius $R_{\rm L}$
are considered, $R_{\rm L}=(0.8,1.0,1.2)$ fm with the diffuseness $a_{\rm L}$
fixed at $a_{\rm L}=R_{\rm L}/2$ in each case. 

The main difference between the potentials constructed in Ref.~\cite{Piarulli15}
and those in the current work
lies in the operator structure of their short-range components, which
we now take to have the form
\begin{equation}
\label{eq:vr}
v_{12}^{\rm S}=\sum_{l=1}^{16} v_{\rm S}^l (r) \, O^l_{12}\ ,
\end{equation}
where $O^{l=1,\dots,6}_{12}$ have been defined above,
\begin{equation}
O^{l=7,\dots,11}_{12}={\bf L}\cdot{\bf S}\,,\,
 {\bf L}\cdot{\bf S}\,{\bm \tau}_1\cdot {\bm \tau}_2\, ,\, ({\bf L}\cdot{\bf S})^2\, ,\, {\bf L}^2\, ,\, 
 {\bf L}^2\, {\bm \sigma}_1\cdot {\bm \sigma}_2 \ ,
 \end{equation}
and
\begin{equation}
O^{l=12,\dots,16}_{12}=T_{12}\, , \, \left(\tau_1^z+\tau_2^z\right)
\, , \,{\bm \sigma}_1\cdot {\bm \sigma}_2\, T_{12}\, ,\, S_{12}\,T_{12}\, ,\,
 {\bf L}\cdot{\bf S}\, T_{12} \ .
 \end{equation}
The parametrization above differs in two ways from that of the minimally non-local
potential of Ref.~\cite{Piarulli15}.  The first difference concerns the ${\bf p}^2$ terms
\[
\{\,v_{\rm S}^p(r)+v_{\rm S}^{p\sigma}(r)\,{\bm \sigma}_1\cdot {\bm \sigma}_2\nonumber
+v_{\rm S}^{pt}(r)\,S_{12}+v_{\rm S}^{pt\tau}(r)\,S_{12}\,{\bm \tau}_1\cdot {\bm \tau}_2\,\, ,\,\,{\bf p}^2\,\}\ ,
\]
which are now absent in Eq.~(\ref{eq:vr}).
The second difference has to do with the charge-symmetry breaking (CSB) piece of 
$v_{12}^{\rm S}$, which, in contrast to Ref.~\cite{Piarulli15}, includes only
the LO term proportional to $\left(\tau_1^z+\tau_2^z\right)$ needed
to reproduce the singlet $nn$ scattering length.

The radial functions $v_{\rm S}^l (r)$ are the same
as those listed in Appendix B of Ref.~\cite{Piarulli15}, and involve
a local regulator (to replace the $\delta$ functions) taken as 
\begin{equation}
 C_{R_{\rm S}}(r)=\frac{1}{\pi^{3/2}R_{\rm S}^3} e^{-(r/R_{\rm S})^2} ,
\end{equation}
where we consider, in combination with $R_{\rm L}=(0.8,1.0,1.2)$ fm,
$R_{\rm S}=(0.6,0.7,0.8)$ fm, corresponding to typical momentum-space
cutoffs $\Lambda_{\rm S}=2/R_{\rm S}$ ranging from about 660 MeV down to 500 MeV.
Hereafter we will denote the potential with cutoffs $(R_{\rm L},R_{\rm S})=(1.2,0.8)$ fm
as model $a$, that with $(1.0,0.7)$ fm as model $b$, and that with $(0.8,0.6)$ fm as
model $c$.  These radial functions contain 26 LECs.  Of these, 20 are in the charge-independent
part of $v_{12}^{\rm S}$: 2 at LO ($Q^0$), 7 at NLO ($Q^2$), and 11 at N3LO ($Q^4$).  The remaining 6 are in its charge-dependent part:
2 at LO (one each from CIB and CSB), and 4 at NLO from CIB.
The optimization procedure to fix these 26 LECs is the same as that adopted
in Ref.~\cite{Piarulli15}, and is discussed in the next section.
It uses $pp$ and $np$ scattering data (including normalizations), as assembled
in the Granada database~\cite{Navarro1314}, the $nn$ scattering length,
and the deuteron binding energy. The minimization of the objective function $\chi^2$
with respect to the LECs is carried out with the Practical Optimization Using no Derivatives
(for Squares), POUNDerS~\cite{POUNDerS}.

\section{Total $\chi^2$ and phase shifts}
\label{sec:phases}
\begin{table*}[t]
\begin{center}
\begin{tabular}{ccccc}
\hline\hline
model &  order&  $E_{\rm Lab}$ (MeV) & $N_{pp+np}$& $\chi^2$/datum \\
\hline 
$b$& LO     &  0--125 &2558 &59.88 \\
$b$& NLO   & 0--125 &2648 &2.18 \\
$b$& N2LO  & 0--125 &2641 &2.32 \\
$b$  & N3LO & 0--125 &2665 &1.07 \\
\hline
$a$  & N3LO  & 0--125 &2668& 1.05 \\
$c$& N3LO  & 0--125&2666 & 1.11 \\
 \hline
$\widetilde{a}$  & N3LO &  0--200& 3698 & 1.37\\
$\widetilde{b}$  &  N3LO & 0--200& 3695& 1.37\\ 
$\widetilde{c}$  & N3LO&  0--200 & 3693&1.40\\
\hline
$a$  & N3LO  & 0--200  & 3690& 2.41 \\
$b$  & N3LO   &0--200  & 3679& 3.76\\
$c$  & N3LO  & 0--200  & 3679 & 4.52 \\
\hline\hline
\end{tabular}
\caption{\label{table:chi2}Total $\chi^{2}$/datum for model $a$ ($\tilde{a}$) with $(R_{\rm L},R_{\rm S})=(1.2,0.8)$ fm,
model $b$ ($\tilde{b}$) with $(1.0,0.7)$ fm, and model $c$ ($\tilde{c}$) with $(0.8,0.6)$ fm fitted up to 125 (200) MeV laboratory energy. For model $b$, results of the fits up to 125 MeV order by order in the chiral expansion are also given;
$N_{pp+np}$ denotes the total number of $pp$ and $np$ data, including observables and normalizations.}
\end{center}
\end{table*}
We report results for the local potentials $v_{12}+v_{12}^{\rm EM}$ described in the previous
section and corresponding to three
different choices of cutoffs $(R_{\rm L},R_{\rm S})$: model $a$ with $(1.2,0.8)$ fm,
model $b$ with $(1.0,0.7)$ fm, and model $c$ with $(0.8,0.6)$ fm.
Models $a$, $b$, and $c$ are fitted to the Granada database of $pp$ and $np$ observables
in two different ranges of laboratory energies, either 0--125 MeV or 0--200 MeV,
to the deuteron binding energy and $nn$ singlet scattering length.
For convenience potential models $a$, $b$, and $c$ fitted up to 200 MeV laboratory energy
are labelled as $\widetilde{a}$, $\widetilde{b}$ and $\widetilde{c}$, respectively.
We list the total number of $pp$ and $np$ data (including normalizations) 
and corresponding total $\chi^2$ per datum for all the potentials in Table~\ref{table:chi2}.
The total number of data points, $N_{pp+np}$, changes slightly for each of the various models 
because of fluctuations in the number of normalizations (see Ref.~\cite{Piarulli15} for more details on
the fit procedure).
For model $b$ we performed fits of the Granada database up to 125 MeV order by order in the chiral expansion. The total $\chi^2$/datum are 59.88, 2.18, 2.32 and 1.07 at LO, NLO, N2LO and N3LO, respectively. There is a strong reduction in the total $\chi^2$ going from LO and NLO and from N2LO and N3LO. However, the quality of the fit worsens slightly in going from NLO to N2LO. At N2LO we fixed the
chiral LECs, namely $c_1$, $c_2$, $c_3$, $c_4$ and $b_3+b_8$, from the $\pi N$ scattering analysis
of Ref.~\cite{Krebs07}.
In the range 0--125 MeV, the total $\chi^2$/datum at N3LO are 1.05, 1.07, 1.11 for models
$a$, $b$, and $c$, respectively; while in the range 0--200 MeV the total $\chi^2$/datum at N3LO
are 1.37, 1.37, 1.40. The total $\chi^2$/datum at N3LO for models 
$a$, $b$, and $c$ when compared (without refitting) to the 0--200 MeV database
are 2.41, 3.76, 4.52, respectively.
In both energy ranges, the quality of the fits deteriorates
slightly as the $(R_{\rm L},R_{\rm S})$ cutoffs are reduced from the values
(1.2,0.8) fm of model $a$ down to (0.8,0.6) fm of model $c$.

The fitted values of the LECs corresponding to models $a$, $b$, $c$ and 
$\widetilde{a}$, $\widetilde{b}$, $\widetilde{c}$ are listed in Tables~\ref{tb:lecsv_125}
and~\ref{tb:lecsv_200}, respectively.
The values for the $\pi N$ LECs in the OPE and TPE terms of these models
are given in Table I of Ref.~\cite{Piarulli15}.

\begin{figure*}
\includegraphics[width=4.5in]{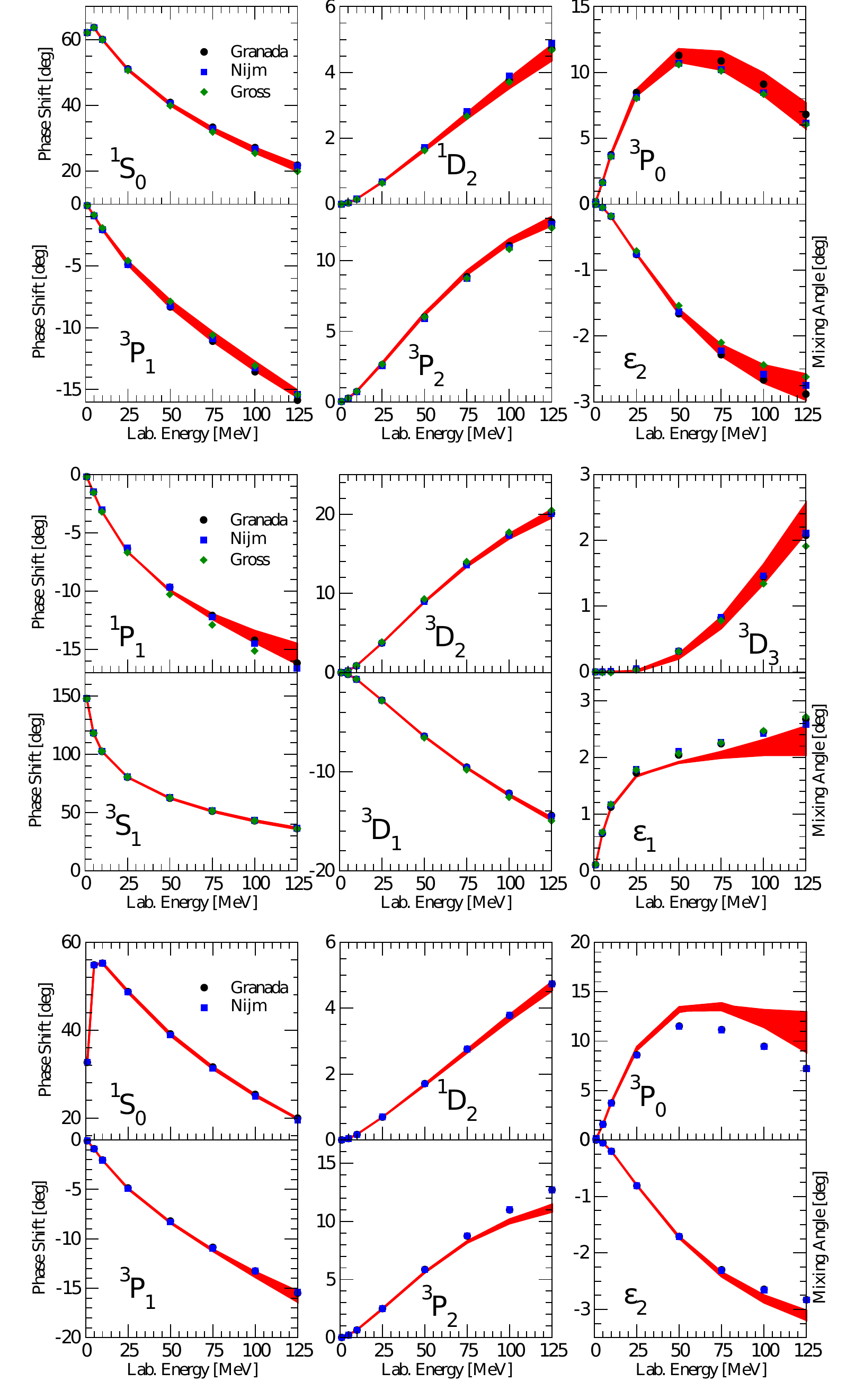}
\caption{(Color online) S-wave, P-wave, and D-wave 
phase shifts for
$np$ in $T$=0 and 1 states (top two panels) and $pp$ (lower panel), obtained
in the Nijmegen~\cite{Stoks93,Stoks94}, Gross and Stadler~\cite{Gross08}, and
Granada~\cite{Navarro1314} PWA's, are compared
to those of models $a$, $b$, and $c$, indicated by the band.}
\label{fig:f1}
\end{figure*}

\begin{figure*}
\includegraphics[width=4.5in]{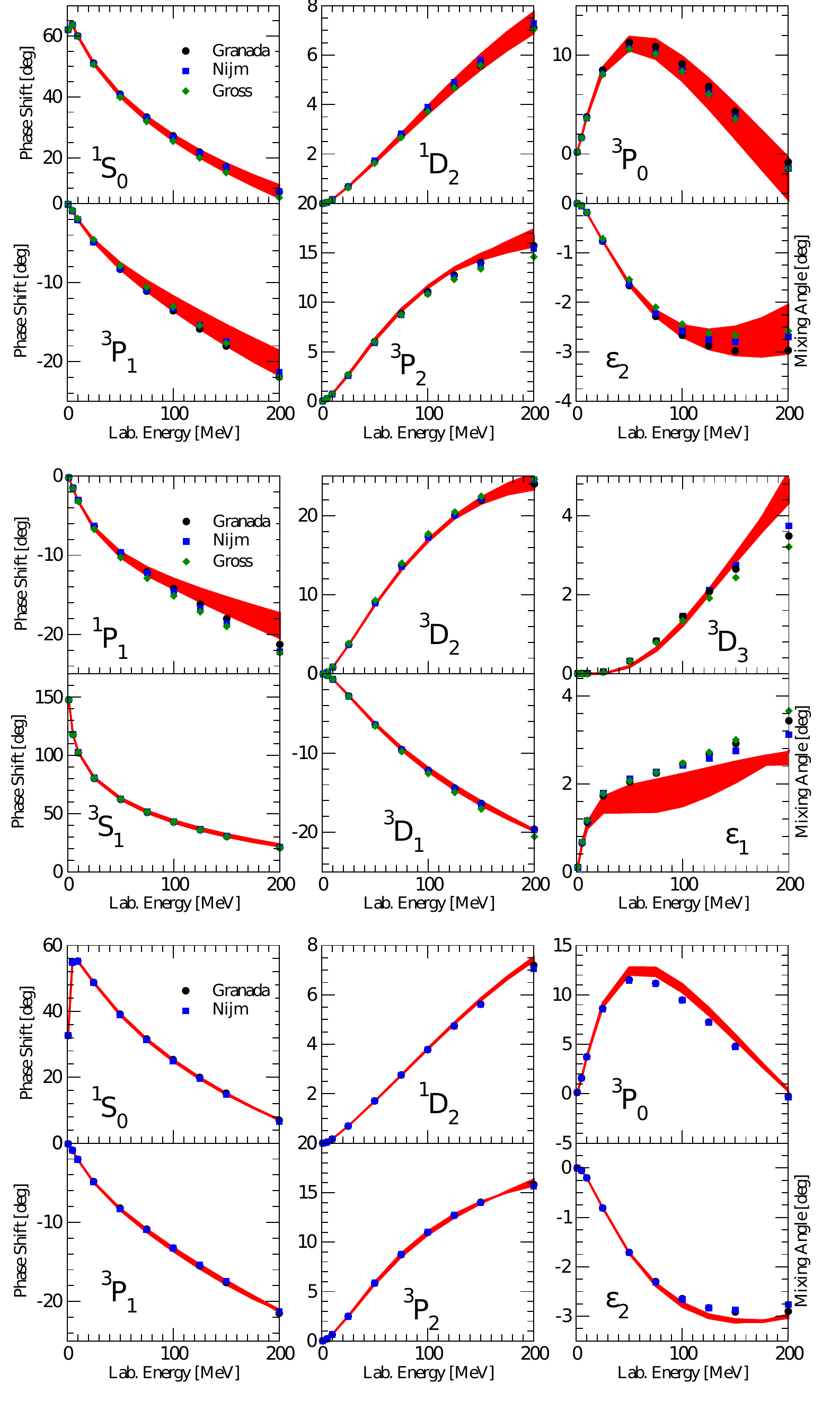}
\caption{(Color online) Same as Fig.~\ref{fig:f1} but for models
$\widetilde{a}$, $\widetilde{b}$, and $\widetilde{c}$ fitted to 200 MeV lab energy.}
\label{fig:f2}
\end{figure*}

The $np$ and  $pp$ S-wave, P-wave, and D-wave phase shifts 
for potential models fitted up to 125 MeV and 200 MeV
laboratory energy are displayed in Figs.~\ref{fig:f1} and~\ref{fig:f2},
respectively. The top two panels of these figures 
show the phase shifts for $np$ in $T\,$=$\,1$ and $T\,$=$\,0$ channels,
respectively, while the remaining bottom panels show the
$pp$ phase shifts (in $T\,$=$\,1$ channel).
The width of the shaded band represents the cutoff sensitivity of the 
phases obtained with the full models $a$, $b$, and $c$, including strong and
electromagnetic interactions.  The calculated phases are compared 
to those obtained in PWA's
by the Nijmegen~\cite{Stoks93}, Granada~\cite{Navarro1314}, and
Gross-Stadler~\cite{Gross08} groups. The recent Gross and Stadler's
PWA is limited to $np$ data only.

\begin{figure*}[bth]
\includegraphics[width=4.5in]{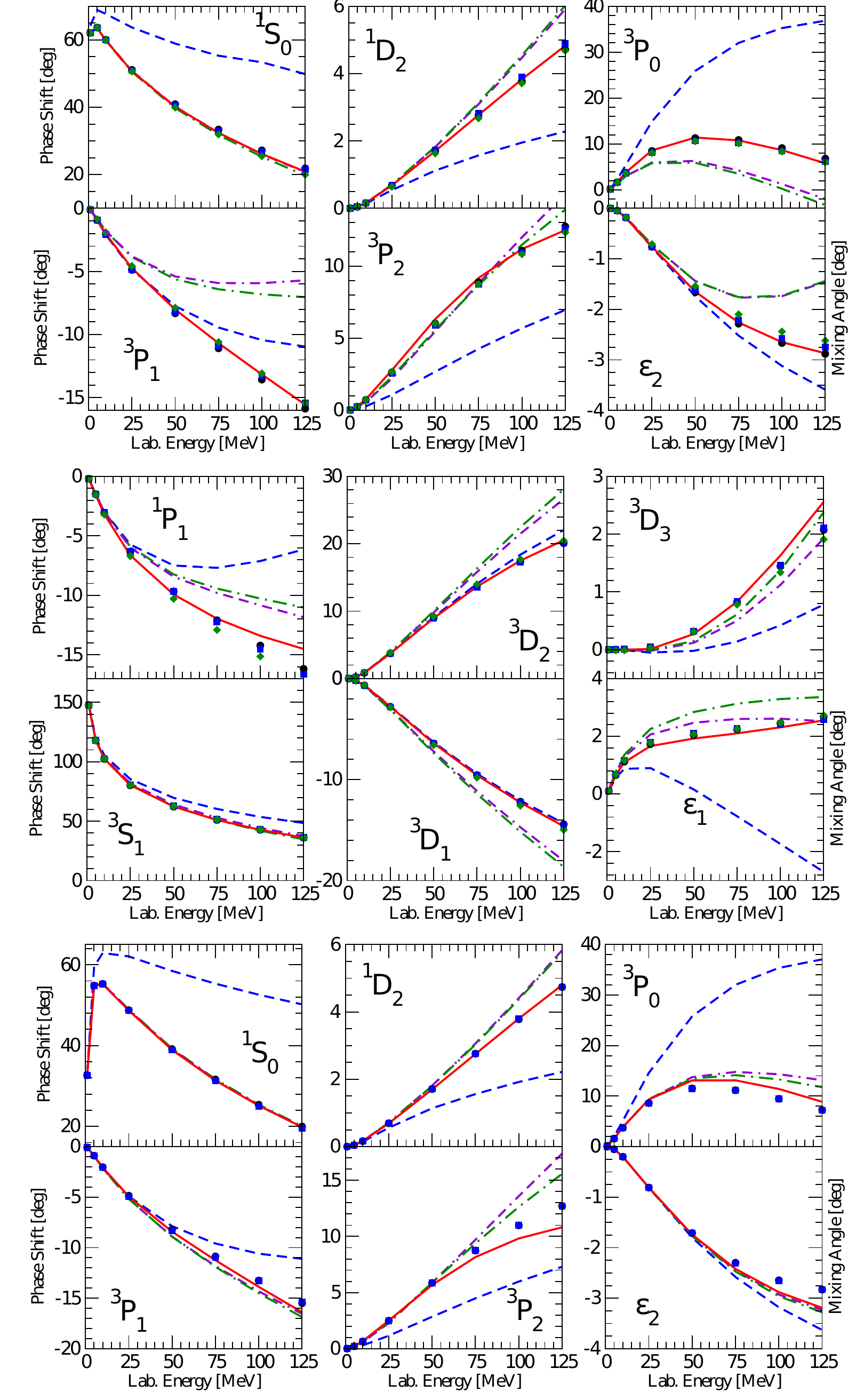}
\caption{(Color online) Chiral expansion of the $np$ (top two panels) and
$pp$ (bottom panel) S-wave, P-wave, and D-wave phase shifts up to 125 MeV 
for model $b$ in comparison with the Nijmegen~\cite{Stoks93,Stoks94}, Gross and Stadler~\cite{Gross08}, and
Granada~\cite{Navarro1314} PWA's. Dashed (blue), dash-dotted (green), double-dash-dotted (magenta), and solid (red) lines show the results at LO, NLO, N2LO and N3LO, respectively.}
\label{fig:f3}
\end{figure*}

In Fig.~\ref{fig:f3}, the $np$ (top panels) and $pp$ (lower panel) S-wave,
P-wave, and D-wave phase shifts are displayed for model $b$ up to 125 MeV
lab energy order-by-order in the chiral expansion. Dashed (blue), dash-dotted
(green), double-dash-dotted (magenta), and solid (red) lines represent the
results at LO, NLO, N2LO and N3LO, respectively. Of course, the description
of the phase shifts improves substantially, as one progresses from LO to N3LO.
\begin{table*}[bth]
\caption{Values of the LECs corresponding to potential models $a$, $b$, $c$ (fitted up to 125 MeV lab energy). The notation $(\pm \,n)$ means $\times 10^{\pm n}$.}
\label{tb:lecsv_125}
\begin{ruledtabular}
\begin{tabular}{lddd}
\textrm{LECs}&
\multicolumn{1}{c}{Model $a$}&
\multicolumn{1}{c}{Model $b$}&
\multicolumn{1}{c}{Model $c$}\\
\colrule
  $C_S$ (fm$^2$) 	& 0.2726141(+1)	& 0.8038124(+1)	&0.1858356(+2)\\
 $C_T$ (fm$^2$) 	&-0.5228448		&-0.1203741(+1)	&-0.6118406(+1)\\
\hline
 $C_1$ (fm$^4$)	&-0.6992838(-1)	&-0.2280422		&-0.5624246\\
 $C_2$ (fm$^4$)	&-0.1496013	 	&-0.2249889 		&-0.3529711\\
 $C_3$ (fm$^4$)	&-0.2502401(-1)	&-0.4007665(-1)	&-0.2225345 \\
 $C_4$ (fm$^4$)	&-0.2728396(-1)	& 0.1243960(-1)	&0.3381613(-1)\\
 $C_5$ (fm$^4$)	&-0.6530008(-2)	&-0.1870727(-1)	&-0.2881762(-1)\\
 $C_6$ (fm$^4$)	&-0.7554924(-1)	&-0.7406609(-1)	&-0.6535759(-1)\\
 $C_7$ (fm$^4$)	&-0.1017206(+1)	&-0.1197452(+1)	&-0.1464748(+1)\\
\hline
 $D_1$ (fm$^6$)	&-0.4251199(-1)	&-0.3820959(-1)	&-0.2163208(-1)\\
 $D_2$ (fm$^6$)	&-0.5567938(-2)	&-0.5343034(-2)	&0.2866318(-2)\\
 $D_3$ (fm$^6$)	&-0.1666607(-1)	&-0.1601394(-1)	&-0.1472287(-1)\\
 $D_4$ (fm$^6$)	&0.1054347(-2)	&0.4219347(-2)	&0.1052796(-2)\\
 $D_5$ (fm$^6$)	&0.5383828(-2)	&0.8971752(-2)	&0.7477159(-2)\\
 $D_6$ (fm$^6$)	&-0.8012050(-2)	&-0.5986245(-2)	&-0.2247046(-2)\\
 $D_7$ (fm$^6$)	&-0.2309392(-1)	&-0.6180197(-2)	&0.3616700(-1)\\
 $D_8$ (fm$^6$)	&0.1383136(-1)	&0.1782567(-1)	&0.2903320(-1)\\
 $D_9$ (fm$^6$)	&0.4797012(-1)	&0.3094851(-1)	&0.9175910(-1)\\
 $D_{10}$ (fm$^6$)	&-0.1156876		&-0.8073891(-1)	&-0.1229688	\\
 $D_{11}$ (fm$^6$)	&-0.1453295(-1)	&-0.1162060(-1)	&-0.2671576(-1)\\
\hline\hline
 $C_0^{\rm IV}$ (fm$^2$)&0.9325477(-2)	&0.1018989(-1)	&0.1357818(-1)\\
 $C_0^{\rm IT}$ (fm$^2$)&0.1578240(-1)	&0.2416591(-1)	&0.2195881(-1)\\
\hline 
 $C_1^{\rm IT}$ (fm$^4$)&-0.2179452(-2)	&-0.3707396(-2)	&-0.2698274(-2)\\
 $C_2^{\rm IT}$ (fm$^4$)&-0.6288540(-2)	&-0.3601899(-2)	&-0.1288174(-2)\\
 $C_3^{\rm IT}$ (fm$^4$)&-0.5799803(-2)	&-0.4559006(-2)	&-0.3126089(-3)\\
 $C_4^{\rm IT}$ (fm$^4$)& 0.2250167(-1)	& 0.1859997(-1)	& 0.8987538(-2)\\
\end{tabular}
\end{ruledtabular}
\end{table*}
\begin{table*}[bth]
\caption{Same as Table~\ref{tb:lecsv_125} but for potential models $\widetilde{a}$, $\widetilde{b}$, $\widetilde{c}$ (fitted up to 200 MeV lab energy).}
\label{tb:lecsv_200}
\begin{ruledtabular}
\begin{tabular}{lddd}
\textrm{LECs}&
\multicolumn{1}{c}{Model $\widetilde{a}$} &
\multicolumn{1}{c}{Model $\widetilde{b}$}&
\multicolumn{1}{c}{Model $\widetilde{c}$}\\
\colrule
  $C_S$ (fm$^2$) 	&0.2936041(+1)	& 0.8398499(+1)	&0.1858331(+2)\\
 $C_T$ (fm$^2$) 	&-0.4933897		&-0.1207696(+1)	&-0.6116424(+1)\\
\hline
 $C_1$ (fm$^4$)	&-0.1013462		&-0.2324413	&-0.5565484\\
 $C_2$ (fm$^4$)	&-0.1444844		&-0.2108143	&-0.3574422\\
 $C_3$ (fm$^4$)	&-0.3647634(-1)	&-0.3461629(-1)	&-0.2266117\\
 $C_4$ (fm$^4$)	&-0.1630825(-1)	& 0.8748772(-2)	&0.3921168(-1)\\
 $C_5$ (fm$^4$)	&-0.6658100(-2)	&-0.3614304(-1)	&-0.2661419(-1)\\
 $C_6$ (fm$^4$)	&-0.6176835(-1)	&-0.5542581(-1)	&-0.6532432(-1) \\
 $C_7$ (fm$^4$)	&-0.9578191		&-0.1019849(+1)	&-0.1465875(+1) \\
\hline
 $D_1$ (fm$^6$)	&-0.3102824(-1)	&-0.1193597(-1)	&-0.2144023(-1)\\
 $D_2$ (fm$^6$)	&-0.4438695(-2)	&-0.4450346(-2)	&0.1386494(-2)\\
 $D_3$ (fm$^6$)	&-0.1351171(-1)	&-0.9542801(-2)	&-0.1620926(-1)\\
 $D_4$ (fm$^6$)	& -0.7084459(-3)	&0.3976205(-2)	&0.2071219(-2) \\
 $D_5$ (fm$^6$)	&  0.1110108(-1)	&0.7809205(-2) 	&0.7238077(-2)\\
 $D_6$ (fm$^6$)	&-0.8598857(-2)	&-0.7362895(-2)	&-0.2323562(-2)\\
 $D_7$ (fm$^6$)	&-0.5367908(-1)	&-0.4158494(-2)	&0.3065351(-1)\\
 $D_8$ (fm$^6$)	&0.3119241(-1)	&0.1090986(-1)	&0.2957488(-1)\\
 $D_9$ (fm$^6$)	&0.3281636(-1) 	&0.6095858(-3)	&0.9135194(-1)\\
 $D_{10}$ (fm$^6$)	&-0.8647128(-1)	&-0.5432144(-1)	&-0.1196465\\
 $D_{11}$ (fm$^6$)	&-0.1167788(-1)	&-0.5186422(-2)	&-0.3065569(-1) \\
\hline\hline
 $C_0^{\rm IV}$ (fm$^2$)&0.9575695(-2)	& 0.1077541(-1)& 0.1312712(-1)\\
 $C_0^{\rm IT}$ (fm$^2$)&0.2194758(-1)	&0.2102140(-1)	&0.1394723(-1)\\
\hline 
 $C_1^{\rm IT}$ (fm$^4$)&-0.1550501(-2)	&0.1152693(-3)	&-0.8965197(-2)\\
 $C_2^{\rm IT}$ (fm$^4$)&-0.8354679(-2)	&-0.1391786(-2)	&-0.3079018(-2)\\
 $C_3^{\rm IT}$ (fm$^4$)&-0.6682746(-2)	&-0.3194459(-3)	&0.3905867(-4)\\
 $C_4^{\rm IT}$ (fm$^4$)&0.1276971(-1)	&0.2879873(-2)	&0.8844043(-3)\\
\end{tabular}
\end{ruledtabular}
\end{table*}
The low-energy scattering parameters are listed in Table~\ref{tab:lep}, where they are compared
to experimental results~\cite{Bergervoet88,Sanden83,Chen08,Miller90,Machleidt01}. The singlet and triplet $np$, and singlet $pp$ and $nn$, scattering lengths
are calculated with the inclusion of electromagnetic interactions.  Without the latter,
the effective range function is simply given by $F(k^2)=k\, \cot \delta=-1/a+r \, k^2/2$ up to terms
linear in $k^2$.  In the presence of electromagnetic interactions, a more complicated
effective range function must be used; it is reported in Appendix D of Ref.~\cite{Piarulli15},
along with the relevant references. 
\begin{table*}[t]
\caption{\label{tab:lep}
The singlet and triplet $np$, and singlet $pp$ and $nn$, scattering lengths and
effective ranges corresponding to the potential models $a$, $b$ and $c$
(fitted up to 125 MeV lab energy), 
and $\widetilde{a}$, $\widetilde{b}$, $\widetilde{c}$ (fitted up to 
200 MeV lab energy). Experimental
values are from Refs.~\cite{Bergervoet88,Sanden83,Chen08,Miller90,Machleidt01}.}
\begin{ruledtabular}
\begin{tabular}{lddddddd}
\textrm{}&
\multicolumn{1}{r}{\textrm{Experiment}}&
\multicolumn{1}{c}{\textrm{Model $a$}}&
\multicolumn{1}{c}{\textrm{Model $b$}}&
\multicolumn{1}{c}{\textrm{Model $c$}}&
\multicolumn{1}{c}{\textrm{Model $\widetilde{a}$}}&
\multicolumn{1}{c}{\textrm{Model $\widetilde{b}$}}&
\multicolumn{1}{c}{\textrm{Model $\widetilde{c}$}}\\
\colrule
$^{1}a_{pp}$	&-7.8063(26)		&-7.776 	 	&-7.774 		&-7.769	&-7.775		&-7.770	&-7.769\\
			&-7.8016(29)		&	 	 	&			&		&			&&\\
$^{1}r_{pp}$	& 2.794(14)   		& 2.780		&2.771		&2.754	&2.774		&2.760	&2.753\\
 			& 2.773(14)   		& 			&			&		&			&		&\\
$^{1}a_{nn}$	&-18.90(40)		&-18.896 		&-18.921		&-18.966	&-18.904		&-19.009	&-18.919\\
$^{1}r_{nn}$	& 2.75(11)  		& 2.825		&2.815		&2.795	& 2.819		&2.801	&2.794\\
$^{1}a_{np}$	&-23.740(20) 		&-23.722		&-23.739		&-23.741	&-23.758		&-23.754 &-23.740\\
$^{1}r_{np}$	& 2.77(5)			& 2.666		&2.686		&2.684	& 2.642		& 2.682	&2.683\\
$^{3}a_{np}$	& 5.419(7)  		& 5.424		&5.424		&5.423	& 5.399		&5.394	&5.424\\
$^{3}r_{np}$	& 1.753(8)   		&1.761 		&1.760		&1.770	&1.727		&1.720	&1.773\\
\end{tabular}
\end{ruledtabular}
\end{table*}

The static deuteron properties are shown in Table~\ref{tab:deut} and compared
to experimental values~\cite{Ericson83,Rodning90,Huber98,Martorell95}.
The binding energy $E_d$ is fitted exactly and includes the contributions (about
20 keV) of electromagnetic interactions, among which the largest is that due to the
magnetic moment term.  The asymptotic S-state normalization, $A_{\rm S}$,
deviates less than 1\% from the experimental data,
and the D/S ratio, $\eta$, is $\sim 2$ standard deviations from experiment
for all models considered.  The deuteron (matter) radius, $r_d$, is
under-predicted by about 0.2$-$1.0\%. It should be noted that this observable has negligible
contributions due to two-body electromagnetic operators~\cite{Piarulli13}.  The
magnetic moment, $\mu_d$, and quadrupole moment, $Q_d$, experimental values
are underestimated by all models, but these observables are known to have significant
corrections from (isoscalar) two-body terms in nuclear electromagnetic charge
and current operators~\cite{Piarulli13}.  Their inclusion would bring the calculated values
considerably closer to experiment.
\begin{table}[bth]
\caption{\label{tab:deut}%
Same as in Table~\protect\ref{tab:lep} but for the deuteron static properties; experimental
values are from Refs.~\cite{Ericson83,Rodning90,Huber98,Martorell95}.}
\begin{ruledtabular}
\begin{tabular}{llllllll}
\textrm{}&
{\textrm{Experiment}}&
{\textrm{Model $a$}}&
{\textrm{Model $b$}}&
{\textrm{Model $c$}}&
{\textrm{Model $\widetilde{a}$}}&
{\textrm{Model $\widetilde{b}$}}&
{\textrm{Model $\widetilde{c}$}}\\
\colrule
$E_{d}$ (MeV)			&2.224575(9)	&2.224574	&2.224573	&2.224576	&2.224574	&2.224568	&2.224570\\
$A_{\rm S}$(fm$^{-1/2}$) & 0.8846(9)	&0.8862		&0.8861   	&0.8874		&0.8811  		&0.8799   	&0.8877     \\
$\eta$				&0.0256(4)	&0.0249		&0.0248		&0.0250		&0.0247		&0.0245		&0.0250	\\
$r_{d}$ (fm)			&1.97535(85)	&1.968		&1.968		&1.971		&1.956		&1.955		&1.971	\\
$\mu_{d}$ ($\mu_{0}$)	&0.857406(1)	&0.850		&0.849		&0.850		&0.850 		&0.850		&0.849	\\
$Q_{d}$	 (fm$^2$)		&0.2859(3)	&0.268		&0.267		&0.269		&0.263		&0.256		&0.269	\\
$P_{d}$	(\%)			&			&5.24		&5.49		&5.32		&5.22		&5.21		&5.35 	\\
\end{tabular}
\end{ruledtabular}
\end{table}
\section{HH and QMC calculations for light nuclei}
\label{sec:nuclei}
The study of light nuclei is especially interesting since it provides the opportunity
to test, in essentially exact numerical calculations, models of two- and three-nucleon forces.
In this section, we briefly discuss the HH and QMC methods adopted here for the accurate or exact solution of 
the few-nucleon Schr\"{o}dinger equation, $H\,\Psi=E\, \Psi$, where $\Psi$ is a nuclear wave function
with specific spin, parity and isospin.   We then present results for the binding energies
and rms radii of the $A\,$=$\,2$--6 nuclei with a Hamiltonian $H$ including the nonrelativistic
kinetic energy in combination with the two-body potentials $v_{12}$ of Sec.~\ref{sec:pots}.
In particular for our calculations we use nuclear wave functions corresponding to
models $a$, $\widetilde{a}$ and $b$, $\widetilde{b}$, whose LECs are specified in
Tables~\ref{tb:lecsv_125} and~~\ref{tb:lecsv_200}.

The HH method is used to calculate the ground-state energies of $^3$H and
$^4$He and these results provide a benchmark for the corresponding QMC
calculations.  The QMC methods are then applied to compute binding energies and rms radii 
of the $^3$He ground state, of the $^6$Li and $^6$He ground and excited states.

\subsection{ The Hyperspherical Harmonics Method}
The HH method uses hyperspherical-harmonics functions as a suitable 
expansion basis for the wave function of an $A$-body system. In the specific case of 
$A\,$=$\,3$ and 4 nuclei, the corresponding ground-state wave functions
 $\Psi^{J^\pi}_A$ ($J^\pi$ being the total angular momentum and parity) 
 can be expanded in the following way:
 \begin{equation}
  \Psi_3^{1/2^+}=\sum_{[K_3]} u_{[K_3]}(\rho_3) {\cal B}_{[K_3]}(\Omega_3)\ ,
\label{eq1}
 \end{equation}
and 
 \begin{equation}
  \Psi_4^{0^+}=\sum_{[K_4]} u_{[K_4]}(\rho_4) {\cal B}_{[K_4]}(\Omega_4)\ .
\label{eq2}
 \end{equation}
Here ${\cal B}_{[K_3]}(\Omega_3)$ and ${\cal B}_{[K_4]}(\Omega_4)$ are fully
antisymmetrized HH-spin-isospin functions for three and four nucleons 
characterized by the set of
quantum numbers $[K_3]\equiv [n_1,l_1,l_2,L,s,S,t,T]$ and
$[K_4]\equiv [n_1,n_2,l_1,l_2,l_3,l',L,s,s',S,t,t',T]$
respectively. 
The quantum numbers $n_i,l_i$ and $l'$ enter in the construction of the
HH vector and are such that the grand angular momenta are 
$K_3=2n_1+l_1+l_2$ and $K_4=2n_1+2n_2+l_1+l_2+l_3$.
The orbital angular momenta $l_i$ (and $l'$ for $A=4$) are
coupled to give the total orbital angular momentum $L$. The
total spin and isospin of the vector are indicated with $S$ and $T$, 
respectively, and $s,s',t,t'$ are intermediate couplings.
A detailed description of the HH method with the
explicit expression of the HH-spin-isospin functions can be found in 
Refs.~\cite{Kie94,Kie97,Viv05,Kie08}.

The hyperspherical coordinates $(\rho_A,\Omega_A)$ in Eqs.~(\ref{eq1}) 
and~(\ref{eq2}) are given by the hyperradius, 
$\rho_A^2=\sum_{i=1}^{A-1} {\bf x}_i^2$
expressed in terms of the $A$--$1$ Jacobi vectors ${\bf x}_i$ of the systems, 
and the hyperangles 
$\Omega_A=({\hat{\bf x}}_1 \ldots {\hat{\bf x}}_{A-1},\alpha_2\ldots\alpha_{A-1})$, with
${\hat{\bf{x}}}_i$ being the unit Jacobi vectors 
and $\alpha_i$ the hyperangular variables.
For $A=3$, $\cos\alpha_2=x_2/\rho_3$, and for $A=4$ 
$\cos\alpha_2=x_2/\sqrt{x_1^2+x_2^2}$ and $\cos\alpha_3=x_3/\rho_4$~\cite{Kie08}.

In the present application of the HH method, the hyperradial functions 
are in turn
expanded in terms of generalized Laguerre polynomials multiplied by
an exponential function
\begin{equation}
u_\mu(\rho_A)= \sum_m C_{m,\mu}\;{\cal L}_m^{(3A-4)}(z) \; e^{-z/2}\ ,
\label{eq3}
\end{equation}
with $z=\beta\rho_A$, $\beta$ being a nonlinear parameter, and 
$\mu\equiv[K_A]$. Introducing the above expansion in Eqs.~(\ref{eq1}) 
and~(\ref{eq2}), we can rewrite $\Psi_A^{J^\pi}$ in the compact form
 \begin{equation}
  \Psi_A^{J^\pi}=\sum_{m,\mu} C_{m,\mu}\; \Phi_{m,\mu}(\rho_A,\Omega_A) \ ,
\label{eqA}
 \end{equation}
where the (normalized) complete antisymmetric vectors are
\begin{equation}
\Phi_{m,\mu}(\rho_A,\Omega_A) 
= {\cal L}_m^{(3A-4)}(z) e^{-z/2} {\cal B}_{[K_A]}(\Omega_A).
\end{equation} 

The ground state energy $E$ is obtained by applying the Rayleigh-Ritz variational
principle, which leads to the following eigenvalue-eigenstate problem
 \begin{equation}
  \sum_{m',\mu'} (H_{m\mu,m'\mu'}-E I_{m\mu,m'\mu'})=0
\label{eqH}
 \end{equation}
where $H_{m\mu,m'\mu'}$ are the Hamiltonian matrix elements 
$\langle m\mu|H|m'\mu'\rangle$
and $I_{m\mu,m'\mu'}$ indicates the matrix elements of the identity matrix. 
The convergence
of the energy $E$ in terms of the size of the basis is studied as follows. 
The HH functions
are collected in channels having specific combinations of the HH-spin-isospin
quantum numbers.
For the three-nucleon system the basis includes all possible combinations of 
HH functions up to $l_1+l_2=6$ corresponding to $23$ angular-spin-isospin 
channels with isospin components $T=1/2,3/2$.
For each channel the hyperangular quantum number $n_1$ and hyperradial 
quantum number $m$ are increased until
convergence is reached at a level of accuracy of the order of a few
keV on the sought energy eigenvalue. In the case of $A\,$=$\,4$
all possible combinations of HH functions up to
$l_1+l_2+l_3=6$ having $T=0$ are included, while for the wave function components having
$T>0$ HH-spin-isospin states
up to $l_1+l_2+l_3=2$ are considered. This selection corresponds to about $234$ 
angular-spin-isospin channels. For each channel the
hyperangular quantum numbers $n_1,n_2$ and hyperradial quantum number $m$ are increased until
convergence is reached at a satisfactory level of accuracy.  Detailed studies of the convergence
have been done in Ref.~\cite{Viv05}, showing that with this kind of expansion an accuracy of about
20 keV can be obtained for the $^4{\rm He}$ ground state energy.
\subsection{Quantum Monte Carlo Methods}
Over the last three decades, QMC methods have been successfully used to study 
the structure and reactions of light nuclei and nucleonic matter starting from
phenomenological interactions.  The extensive use of these {\it ab-initio} methods
for computing many of the important properties of light nuclei, such as spectra,
form factors, radiative and weak transitions, low-energy scattering and electroweak
response, has led to a rather large number of references, where detailed descriptions of 
QMC algorithms, as well as tests of their accuracy, have been described in detail and
discussed at length (see, for example, the review article~\cite{Carlson14} and references
therein for a complete overview of the topic).  In this section we briefly outline those
features of QMC techniques relevant for the implementation of these methods with
the present chiral (and local) $NN$ potentials at N3LO.

The QMC calculations proceed in two steps.  The first step is the variational Monte
Carlo (VMC) calculation, in which trial wave functions are optimized by minimizing
the Hamiltonian.  The second consists of the Green's function Monte Carlo (GFMC)
calculation, in which the exact wave functions of the nuclear Hamiltonian are projected
out of these optimized trial wave functions by evolving them in imaginary time.

In VMC calculations, one assumes a suitably parametrized form for the antisymmetric
wave function $\Psi_T$ of a given spin, parity and isospin and optimizes the variational
parameters by minimizing the energy expectation value, $E_T$,
\begin{equation}
\label{eq:energy_vmc}
 E_T=\frac{\langle \Psi_T|H|\Psi_T\rangle}{\langle \Psi_T|\Psi_T\rangle}\geq E_0 \ ,
\end{equation}
which is evaluated by Metropolis Monte Carlo integration~\cite{Metropolis53}. 
The lowest value for $E_T$ is then taken as the approximate ground-state energy.
Upper bounds to energies of excited states can also be obtained, either from standard VMC calculations
if they have different quantum numbers from the ground state, or from small-basis
diagonalizations if they have the same quantum numbers.

The ``best'' variational wave functions $\Psi_T$ for the nuclei studied in the present
work have the form~\cite{Wiringa91}
\begin{equation}
 |\Psi_T\rangle =S\,\prod_{i<j}^A\,(1+U_{ij})\,|\Psi_J\rangle \ ,
\end{equation}
where $S$ is the symmetrization operator. The Jastrow wave function $\Psi_J$
is fully antisymmetric and has the ($J^{\pi}; T$) quantum numbers of the state
of interest, while $U_{ij}$ are the two-body correlation operators. The correlation 
functions in $U_{ij}$ are obtained by solving two-body Euler-Lagrange equations
projected in pair spin $S$ and isospin $T$ channels, and for finite nuclei are
required to satisfy suitable boundary conditions~\cite{Wiringa91}.  Since
the calculations carried out here are with only two-body interactions,
three-body correlations induced by three-body interactions are not explicitly
accounted for in $\Psi_T$.  

In order to find the optimum $\Psi_T$, the minimization
of the energy expectation value and its associated variance
are carried out with respect to the variational parameters.
In the case of $A\,$=$\,6$ nuclei, the optimization of the energies is subject to
the constraint that the rms radii are close to the GFMC ones obtained with the AV18.
This is because the best variational wave functions we have 
do not make p-shell nuclei stable against breakup into sub-clusters.
The search for the best sets of variational parameters is performed by using 
the optimization tool NLopt~\cite{NLopt}, a free open-source library for 
nonlinear optimization problems.

Given the best set of variational parameters, the trial wave function $\Psi_T$
can then be used as the starting point of a GFMC~\cite{Carlson87,Carlson88}
calculation which projects out of it the exact lowest energy state $\Psi_0$ with
the same quantum numbers.  The projection of $\Psi_0$ is carried out by evolving
for long imaginary time $\tau=-i\, t$
\begin{equation}
|\Psi_0\rangle \propto \lim_{\tau\to\infty}|\Psi(\tau)\rangle=\lim_{\tau\to\infty}e^{-(H-E_0)\,\tau}\,|\Psi_T\rangle\ ,
\end{equation}
with the obvious initial condition $|\Psi(\tau\!\!=\!\!0)\rangle=| \Psi_T\rangle$.  In practice
the imaginary-time evolution operator ${\rm exp}[-(H-E_0)\,\tau]$ is computed for small time
steps $\Delta \tau$ with $\tau\,$=$\,n\,\Delta \tau$, and 
is carried out with a simplified version $H^{\prime}$ of the full Hamiltonian $H$.
In the presence of only $NN$ interactions the Hamiltonian $H^{\prime}$ contains a
charge-independent eight-operator projection, $\left[{\bf 1}\, ,\, {\bm \sigma}_1\cdot {\bm \sigma}_2\, , \,S_{12}\, ,
{\bf L}\cdot {\bf S}\right] \otimes\left[{\bf 1}\, ,\, {\bm \tau}_1\cdot {\bm \tau}_2\right]$, of the full
two-body potential, constructed to preserve the potential in all S and P waves as well as 
the $^3$D$_1$ and its coupling to the $^3$S$_1$.

The desired expectation values of 
ground-state and low-lying excited-state observables are then
computed approximately by stochastic integration of ``mixed'' matrix elements~\cite{Pudliner97}
\begin{equation}
\label{eq:mme}
\langle{\mathcal O}(\tau)\rangle_{\rm M}= \frac{\langle\Psi(\tau)|{\mathcal O}|\Psi_T\rangle}{\langle\Psi(\tau)|\Psi_T\rangle}\, ,
\end{equation}
where ${\mathcal O}$ is the observable of interest to be evaluated.
By writing $\Psi(\tau)=\Psi_T+\delta \Psi(\tau)$ and neglecting terms of order $[\delta \Psi(\tau)]^2$, 
one obtains an approximate expression for
\begin{equation}
\label{eq:extrap}
 \langle{\mathcal O}(\tau)\rangle\equiv\frac{\langle\Psi(\tau)|{\mathcal O}|\Psi(\tau)\rangle}{\langle\Psi(\tau)|\Psi(\tau)\rangle}\approx \langle{\mathcal O}(\tau)\rangle_{\rm M} +
[\langle{\mathcal O}(\tau)\rangle_{\rm M}-\langle{\mathcal O}\rangle_{\rm V}]\ ,
\end{equation}
where $\langle{\mathcal O}\rangle_{\rm V}$ is the variational expectation value.

In the case of the Hamiltonian, since the propagator commutes with it, the mixed
estimate $\langle H(\tau)\rangle_{\rm M}$ of Eq.~(\ref{eq:mme}) is itself an upper
bound to the the ground-state energy $E_0$ and can be expressed as~\cite{Pudliner97}
\begin{equation}
\label{eq:mmee}
 E(\tau)=\langle H(\tau)\rangle_{\rm M}=\frac{\langle\Psi(\tau/2)|H|\Psi(\tau/2\rangle}{\langle\Psi(\tau/2)|\Psi(\tau/2)\rangle}\ .
\end{equation}
Because the simpler $H^{\prime}$ is used to generate the GFMC propagator
the total energy is then computed by the mixed estimate of $H^{\prime}$ plus the difference
$\langle H-H^{\prime} \rangle_{\rm M}$ evaluated by Eq.~(\ref{eq:extrap}).

Apart from the use of mixed estimates and $H^{\prime}$ in the propagation,
another source of systematic errors that affects GFMC calculations is the
well-known fermion sign problem.  In essence this results from the fact that 
during the imaginary-time propagation bosonic noise gets mixed into
the propagated wave function.  This bosonic component has a much lower
energy than the fermion component and thus is exponentially amplified in
subsequent iterations of the short-time propagators.
The desired fermionic component is projected out by the antisymmetric
$\Psi_T$ when Eq.~(\ref{eq:mme}) is evaluated; however, the presence of large
statistical errors which increase with $\tau$ effectively limits the maximum $\tau$
that can be used in the calculations.  Since the number of pairs to be exchanged
grows with the mass number $A$, the sign problem also grows exponentially
with increasing $A$.

For spin- and isospin-dependent wave functions, the fermion sign problem
can be controlled by a suitable constrained path approximation, which basically
limits the initial propagation to regions where the propagated $|\Psi(\tau)\rangle$
and trial $|\Psi_T\rangle$ wave functions have a positive overlap and discards
those configurations that instead have a small or vanishing overlap
(see Ref.~\cite{Wiringa00} for details on this topic). To address the possible bias
that the constrained path technique can introduce in the calculations, all the
configurations (also those that would be rejected) for a small number of unconstrained
time steps $n_{\rm uc}$ are used when evaluating the expectation values.  In general
the number $n_{\rm uc}$ is chosen to be as large as possible within a reasonable
statistical error.

For phenomenological nuclear Hamiltonians (such those based on the
AV18 potential) the constrained-path approximation was not necessary for
calculations of $A\leq4$ systems, since the sign problem was quite mild for
these light nuclei.  On the other hand, it is essential for GFMC calculations
with the N3LO $NN$ chiral interactions of Sec.~\ref{sec:pots}, since the sign
problem is far more severe for this category of potentials.
\subsection{Results for binding energies}
In this section we present results for ground and excited states of
$^3$H, $^3$He, $^4$He, $^6$He, and $^6$Li nuclei using a subset of the 
local chiral potentials discussed in Sec.~\ref{sec:pots}.  In particular, in order 
to solve the $^3$H and $^4$He ground states, we use VMC, GFMC, and HH methods
with N3LO $NN$ models $a$, $\widetilde{a}$, $b$ and $\widetilde{b}$, while
for $^3$He, $^6$He, and $^6$Li ground and excited states we present VMC
and GFMC calculations performed with model $\widetilde{b}$ only.

The variational wave functions used for the VMC results include 
only spatial and spin-isospin two-body correlations denoted 
by $U_{ij}$ as in Refs.~\cite{Wiringa91,Pudliner97}; the Jastrow
wave functions for the s-shell ($A\,$=$\,3$ and 4) and p-shell ($A\,$=$\,6$)
nuclei are also given explicitly in those references.
For these calculations, the search in parameter space is made using 
COBYLA (Constrained Optimization BY Linear Approximations) algorithm available in
the NLopt~\cite{NLopt} library. The optimal parameters are found typically 
using runs of 100,000 configurations for the evaluation
of matrix elements in Eq.~(\ref{eq:energy_vmc}).
When the optimal trial wave function is found, a long run with 
1,000,000, 500,000, and 200,000 configurations is made 
in $A\,$=$\,3$, 4 and 6 nuclei, respectively, which then is used 
as input for the GFMC calculations.
The GFMC results are obtained using the constrained path 
technique with $n_{\rm uc}\,$=$\,20$ unconstrained time steps. 
The imaginary-time evolution for the $a$ and $\widetilde{b}$ 
models ($\widetilde{a}$ and $b$ ones)
is computed with small time step $\Delta \tau\,$=$\,0.0005\,(0.0001)$
MeV$^{-1}$ up to total time $\tau=0.2$ MeV$^{-1}$.

The results for the $^3$H and $^4$He ground states are shown in 
Tables~\ref{tab:h3} and~\ref{tab:he4}, respectively.  The VMC
calculations give energies that are 3--4\% above the corresponding
HH or GFMC predictions; the latter are in good agreement with
each other.  The errors quoted for the VMC and GFMC results
are the Monte Carlo statistical errors.  We see that increasing the
laboratory energy range, in which the LECs are fitted, from 125 to 200 MeV
(as discussed in Sec.~\ref{sec:phases}), leads to more binding for these systems.

\begin{table}[bth]
\begin{center}
\begin{tabular}{ccccccccc}
\hline\hline
      \multicolumn{1}{c}{} &
      \multicolumn{2}{c}{Model $a$} &
      \multicolumn{2}{c}{Model $\widetilde{a}$} &
      \multicolumn{2}{c}{Model $b$} &
      \multicolumn{2}{c}{Model $\widetilde{b}$}\\
\hline
Method& $E_0$ &  $\sqrt{\langle r_p^2\rangle}$&  $E_0$ &  $\sqrt{\langle r_p^2\rangle}$&$E_0$ &  $\sqrt{\langle r_p^2\rangle}$ &$E_0$&  $\sqrt{\langle r_p^2\rangle}$  \\
\hline
VMC	&--7.592(6)&1.65	&--7.691(6)  &1.62		&--7.317(7)&1.68			&--7.643(5)&1.63	\\
GFMC	&--7.818(8)&1.62	&--7.917(10)&1.60 	&--7.627(17)&1.65  		&--7.863(8)&1.57	\\	
HH		&--7.818 &		&--7.949 &			&--7.599 &				&--7.866 &	\\
\hline\hline
\end{tabular}
\caption{ \label{tab:h3}
The $^3$H ground-state energies $E_0$ (MeV) and rms proton radii $r_p$ (fm)
with models $a$, $\widetilde{a}$, $b$, and $\widetilde{b}$.  Statistical errors
on the energy evaluations are indicated in parentheses for the VMC and GFMC calculations.}
\end{center}
\end{table}
\begin{table}[bth]
\begin{center}
\begin{tabular}{ccccccccc}
\hline\hline
      \multicolumn{1}{c}{} &
      \multicolumn{2}{c}{Model $a$} &
      \multicolumn{2}{c}{Model $\widetilde{a}$} &
      \multicolumn{2}{c}{Model $b$} &
      \multicolumn{2}{c}{Model $\widetilde{b}$}\\
\hline
Method& $E_0$ &  $\sqrt{\langle r_p^2\rangle}$& $E_0$& $\sqrt{\langle r_p^2\rangle}$& $E_0$ &$\sqrt{\langle r_p^2\rangle}$& $E_0$ &$\sqrt{\langle r_p^2\rangle}$  \\
\hline
VMC	&--24.38(1)	&1.51	&--25.03(1)	&1.49	&--22.89(2)	&1.54	&--24.46(2)	&1.49\\
GFMC	&--25.13(5)	&1.49	&--25.71(3)	&1.50	&--23.88(5)	&1.53	&--25.21(4)	&1.45\\	
HH		&--25.15 	&		&--25.80  & 		&--23.96 &		&--25.28	   & \\
\hline\hline
\end{tabular}
\caption{ \label{tab:he4}
Same as in Table~\ref{tab:h3} but for the $^4$He ground state.}
\end{center}
\end{table}

In Table~\ref{table:energy} we report VMC and GFMC calculations 
for $^3$H, $\,^3$He, $^4$He, $^6$He, and $\,^6$Li ground and excited states 
obtained using model $\widetilde{b}$, which has, among the N3LO 
local potentials presented in Sec.~\ref{sec:pots}, the ``best'' behavior 
in terms of sign problem. In that table we also report the corresponding
GFMC calculation obtained with the AV18. We note that for $A\,=\, 3$, 4 and 6
the binding energies obtained using model $\widetilde{b}$ differ by about
0.2\,--\,0.3 MeV, 1.07 MeV, and 1.3\,--\,0.5 MeV, respectively, 
from the corresponding ones obtained using AV18.

The optimization of the $^3$He ground state has been performed using as starting point 
the variational parameters for $^3$H, but varying only the separation energies
and tensor/central ratios---these parameters characterize the asymptotic boundary
conditions imposed on the pair-correlation functions~\cite{Wiringa91}.
The calculated VMC energy, as shown in Table~\ref{table:energy}, is
$\sim 0.2$ MeV above the GFMC one.

The ground state of $^6$He, not bound with respect to the
$^4$He threshold, is a $(J^{\pi},T)\,$=$\,(0^{+}; 1)$ state which 
has predominantly a $^{2S+1}L[n]= {^1}{\rm S}[2]$ 
character (we use spectroscopic notation to denote 
the orbital angular momentum $L$, the spin $S$ and the 
Young diagram spatial symmetry $[n]$ of the state). The 
$(2^{+}; 1)$ first excited state, mostly a ${^1}{\rm D}[2]$ 
state, is above the threshold for decay to $\alpha +2n$ with a 
width of $\approx 100$ keV and we treat it as a stable state.
For both states we allow a possible ${^3}{\rm P}[11]$ admixture 
in the total wave function, and then use generalized eigenvalue 
routines to diagonalize the resulting 2$\times$2 matrix for 
each of them and extract the corresponding 
contributions, ${^1}{\rm S}[2]$ and ${^3}{\rm P}[11]$ 
for the $(0^{+}; 1)$ ground state, and $^1{\rm D}[2]$ 
and ${^3}{\rm P}[11]$ for the $(2^{+}; 1)$ excited state. 
We do not report the calculated energies for the three ${^3}{\rm P}[11]$
states with $(J^{\pi},T)\,$=$\,(2^{+}; 1)$, $(1^{+}; 1)$, and $(0^{+}; 1)$ 
since they have yet to be identified experimentally.

The $p-$shell spectrum for $^6$Li consists of a $(1^{+}; 0)$ ground 
state which is mostly a $^3{\rm S}[2]$ state, a triplet 
of ${^3}{\rm D}[2]$ excited states with $(3^{+}; 0)$, $(2^{+}; 0)$,
and $(1^{+}; 0)$ components, and a singlet of $^1{\rm P}[11]$ 
excited state with a $(1^{+}; 0)$ component, the latter not 
yet identified experimentally. The $^6$Li ground state 
is stable while the excited states are above the $\alpha +d$ threshold,
but we will treat them as bound states below. In addition there are 
$(0^{+}; 1)$ and $(2^{+}; 1)$ excited states that are the isobaric
analogs of the $^6$He states, but they will not be 
discussed here. For the $(1^{+}; 0)$ ground and 
excited states we allow admixtures of $^3{\rm S}[2]$, $^3{\rm D}[2]$
and $^1{\rm P}[11]$ components in the total wave function and 
then diagonalize a 3$\times$3 matrix to extract the 
corresponding contributions. This diagonalization procedure is not necessary for the 
$(3^{+}; 0)$ and $(2^{+}; 0)$ excited states since both of them are pure 
${^3}{\rm D}[2]$ states.
The energies of the ${^3}{\rm D}[2]$ triplet give a measure of the 
effective one-body spin-orbit splitting.
The $J$-averaged centroids for both model $\widetilde{b}$ and AV18 are
3.6 MeV above their respective ground states; 
however the spread between lowest and highest triplet members is 1.5 MeV 
for model $\widetilde{b}$ and 2.1 MeV for AV18.

The minimization of the energy for the $^6$Li ground
state has been carried out by requiring the resulting proton 
rms radius, $r_p$, to be close to the GFMC one obtained with the AV18.
For the excited states, we minimize their energies by requiring that these
excited states have radii larger than the ground state.
A similar optimization strategy has been adopted
for the $^6$He ground and excited states, except that we use
as starting point the $^6$Li variational parameters and
vary only those parameters associated with the single-particle radial functions, $\phi_p$,
in the Jastrow part of the trial wave function~\cite{Pudliner97}.

\begin{table*}
\caption{The $^3$H, $\,^3$He, $^4$He, $^6$He, and $\,^6$Li ground- and excited-state energies in MeV
and proton rms radii $r_p$ in fm with model $\widetilde{b}$ compared with the corresponding GFMC results obtained with the AV18. Statistical
errors on the energy evaluations are indicated in parentheses.}
\begin{ruledtabular}
\begin{tabular}{lcccccc}
\multicolumn{1}{c}{} &
\multicolumn{2}{c}{VMC} &
\multicolumn{2}{c}{GFMC} &
\multicolumn{2}{c}{GFMC(AV18)} \\
$^AZ(J^{\pi};T)$ & $E_0$  &$\sqrt{\langle r_p^2\rangle}$  &   $E_0$    &$\sqrt{\langle r_p^2\rangle}$ &$E_0$    &$\sqrt{\langle r_p^2\rangle}$    \\
\hline
$^3$H$(\frac{1}{2}^+;\frac{1}{2})$	&--7.643(5)	&1.63	&--7.863(8)	&1.57	&--7.610(5)	&1.66\\
$^3$He$(\frac{1}{2}^+;\frac{1}{2})$	&--6.907(5)	&1.84	&--7.115(9)	&1.84	&--6.880(5)	&1.85\\
$^4$He$(0^+;0)$				&--24.46(2)	&1.49	&--25.21(4)	&1.45	&--24.14(1)	&1.49\\
$^6$He$(0^+;1)$				&--22.58(3)	&2.05	&--24.53(6)	&2.07(1)	&--23.76(9) 	&2.06(1)\\	
$^6$He$(2^+;1)$				&--20.94(2)	&2.06	&--22.87(6)	&2.18(2)	&--21.85(9)	&2.11(1)\\
$^6$Li$(1^+;0)$				&--25.86(3)  	&2.58 	&--27.71(8)  	&2.62(1)	&--26.87(9)	&2.58(1)\\
$^6$Li$(3^+;0)$     				&--22.73(3)	&2.59	&--24.56(8) 	&2.59(1)	&--24.11(7)	&2.87(1)\\
$^6$Li$(2^+;0)$				&--21.42(3)	&2.61	&--24.04(9)	&2.79(2)	&--22.75(11)	&2.63(1)\\
$^6$Li$(1^+_2;0)$     	&--20.42(3)  	&2.58 	&--23.09(11)  	&2.89(2)	&--21.99(12)	&2.85(3)\\
\end{tabular}
\end{ruledtabular}
\label{table:energy}
\end{table*}

\section{Summary and Conclusions}
\label{sec:conc}
In the present work we have constructed two classes of chiral
potentials at N3LO, which are fully local in configuration space, for use
(primarily) with HH and QMC methods.  The two classes only differ in
the range of lab energies over which the LECs in the contact interactions
have been fitted to the $NN$ database (as assembled by the Granada
group), either 0--125 MeV (models $a$, $b$, and $c$) with $\chi^2$/datum
$\lesssim 1.1$ for a total of about 2700 data points or 0--200 MeV (models $\widetilde{a}$, $\widetilde{b}$,
and $\widetilde{c}\,$) with $\chi^2$/datum $\lesssim 1.4$ for about 3700 data
points (representing an increase of roughly 40\% in the size of the fitted database
relative to the 0--125 MeV case).  Within a given
class, models $a$, $b$, and $c$ (or $\widetilde{a}$, $\widetilde{b}$,
and $\widetilde{c}\,$) have different short-range and
long-range cutoff radii, respectively $R_{\rm L}$ and $R_{\rm S}$:
$(R_{\rm L},R_{\rm S})\,$=$\,(1.2,0.8)$ fm for models
$a$ and $\widetilde{a}$, $(1.0,0.7)$ fm for models $b$ and $\widetilde{b}$,
and $(0.8,0.6)$ fm for models $c$ and $\widetilde{c}$.  The
cutoff radius $R_{\rm L}$ regularizes the long-range part
of the potential, which includes OPE and TPE terms without and
with excitation of intermediate $\Delta$ isobars.  The cutoff radius
$R_{\rm S}$ provides a range to the $\delta$-functions and their derivatives,
which characterize the contact interactions in the short-range part of the
potential.  These contact interactions require a total
of 26 independent LECs, 20 of which occur in the charge-independent (CI)
component and 6 in the charge-dependent (CD) one (5 for central,
tensor and spin-orbit CIB terms, and 1 for a central CSB term).  These
26 LECs are then constrained by the fits above (their values are listed in Tables~\ref{tb:lecsv_125} and
~\ref{tb:lecsv_200}).

A subset of the potentials---$a$, $\widetilde{a}$, $b$, and $\widetilde{b}$---have been
used in HH, VMC, and GFMC calculations of binding energies and proton rms radii
of nuclei with $A\,$=$\,2$--6.  The GFMC calculations are rather challenging owing
to the serious fermion-sign problem associated with these potentials, even for s-shell
nuclei ($^3$H, $\,^3$He, and $^4$He) (this problem becomes especially severe for
models $c$ and $\widetilde{c}$, and they have not been used in the present work).
However, implementation of the constrained-path algorithm in the course of the
imaginary-time propagation substantially reduces the statistical fluctuations in the energy
evaluation, and leads to $^3$H and $^3$He ground-state energies in excellent
agreement with those obtained in the HH calculations.  All
present models, especially $c$ and $\widetilde{c}\,$, have rather strong
spin-orbit, quadratic orbital angular momentum, and quadratic spin-orbit
components, particularly in the $(S,T)\,$=$\,(1,0)$ channel: for internucleon separation
close to zero, they have values of $\sim2800$ MeV, $\sim 200$ MeV, and $\sim 460$ MeV 
respectively, in this channel.
While these components vanish for nucleon pairs in relative S-wave, they do so,
in the course of a GFMC imaginary-time propagation, only by averaging large values of opposite
signs, thus producing large fluctuations.

The models $\widetilde{a}$ and $\widetilde{b}$ produce more binding in $A\,$=$\, 3$
and 4 nuclei than $a$ and $b$; the extra binding of model $\widetilde{b}$ relative to
$b$ amounts to 5\% in $^4$He.  It appears that model $\widetilde{b}$
leads to ground- and excited-state energies of $A\,$=$\,3$--6 nuclei, which are
close to those calculated with AV18.
Clearly, the next stage in the program of studies
of light nuclei structure
with chiral interactions we envision, is the inclusion of a three-nucleon potential.
A chiral version of it at leading order, including $\Delta$-isobar intermediate
states, has been developed, and is currently being constrained by
reproducing observables in the $A\,$=$\,3$ systems.

\section*{Acknowledgments}
Conversations and e-mail exchanges with J.\ Carlson and S.\ Gandolfi are
gratefully acknowledged.
The work of~M.P.,~A.L.,~S.C.P., and~R.B.W has been supported by 
the NUclear Computational Low-Energy Initiative (NUCLEI) SciDAC
project. This research is supported by the U.S.~Department of Energy, Office of Science, 
Office of Nuclear Physics, under contracts DE-AC02-06CH11357 (M.P.,~A.L.,~S.C.P., and~R.B.W.)
and DE-AC05-06OR23177 (R.S.).
This research also used resources provided by Argonne's Laboratory
Computing Resource Center and by the National Energy Research Scientific Computing Center (NERSC).


\begin{thebibliography}{100}
%
%
%
\bibitem{Barrett13}
B.R.\ Barrett, P.\ Navrtil, and J.P.\ Vary,
Progress in Particle and Nuclear Physics {\bf 69} (0), 131 (2013).
%
\bibitem{Jurgenson13}
E.D.\ Jurgenson, P.\ Maris, R.J.\ Furnstahl, P.\ Navrátil, W.E.\ Ormand, and J.P.\ Vary, 
Phys. Rev. C {\bf 87}, 054312 (2013).
%
\bibitem{Hagen14}
G.\ Hagen, T.\ Papenbrock, A.\ Ekstr\"{o}m, K.A.\ Wendt, G.\ Baardsen, S.\ Gandolfi, 
M.\ Hjorth-Jensen, and C.J.\ Horowitz,
Phys.\ Rev.\ C {\bf 89}, 014319 (2014).
%
\bibitem{Hagen14_1}
G.\ Hagen, T.\ Papenbrock, M.\ Hjorth-Jensen, and D.J.\ Dean,
Reports on Progress in Physics {\bf 77} (9), 096302 (2014).
%
\bibitem{Viviani06}
M.\ Viviani, L.E.\ Marcucci, S.\ Rosati, A.\ Kievsky, and L.\ Girlanda,
Few-Body Syst. {\bf 39}, 159 (2006). 
%
\bibitem{Marcucci09}
L.E.\ Marcucci, A.\ Kievsky, L.\ Girlanda, S.\ Rosati, and M.\ Viviani,
Phys. Rev. C {\bf 80}, 034003 (2009).
%
\bibitem{Viviani09}
M.\ Viviani, A.\ Kievsky, L.\ Girlanda, L.E.\ Marcucci, and S.\ Rosati, 
Few-Body Syst. {\bf 45}, 119 (2009).
%
\bibitem{Viviani10}
M.\ Viviani, L.\ Girlanda, A.\ Kievsky, L.E.\ Marcucci, and S.\ Rosati,
EPJ\ Web\ Conf. {\bf 3}, 05011 (2010).
%
\bibitem{Bogner10}
S.\ Bogner, R.\ Furnstahl, and A.\ Schwenk, 
Progress in Particle and Nuclear Physics {\bf 65} (1), 94 (2010).
%
\bibitem{Hergert13}
H.\ Hergert, S.K.\ Bogner, S.\ Binder, A.\ Calci, J.\ Langhammer, R.\ Roth, and A.\ Schwenk,
Phys.\ Rev.\ C {\bf 87}, 034307 (2013).
%
\bibitem{Dickhoff04}
W.\ Dickhoff, and C.\ Barbieri, 
Progress in Particle and Nuclear Physics {\bf 52} (2), 377 (2004).
%
\bibitem{Soma13}
V.\ Som\`{a}, C.\ Barbieri, and T.\ Duguet, 
Phys. Rev. C {\bf 87}, 011303 (2013).
%
\bibitem{Carlson14}
J.\ Carlson {\it et al.},
Rev.\ Mod.\ Phys.\ {\bf 87}, 1067 (2015).
%
%
\bibitem{Weinberg90}
S.\ Weinberg,
Phys.\ Lett.\ {\bf B251}, 288 (1990);
Nucl.\ Phys.\ {\bf B363}, 3 (1991);
Phys.\ Lett.\ {\bf B295}, 114 (1992).
%
\bibitem{Epelbaum09r}
E.\ Epelbaum, H.\ W.\ Hammer, and U.-G.\ Mei\ss{}ner, 
Rev.\ Mod.\ Phys.\ {\bf 81}, 1773 (2009).
%
\bibitem{Entem11} 
R.\ Machleidt and D.R.\ Entem,
Phys.\ Rep.\ {\bf 503}, 1 (2011).
%
\bibitem{Kalantar12}
N.\ Kalantar-Nayestanaki {\it et al.}, 
Rept.\ Prog.\ Phys.\ {\bf 75}, 016301 (2012).
%
\bibitem{Hammer13}
H.\ W.\ Hammer, A.\ Nogga, and A.\ Schwenk, 
Rev.\ Mod.\ Phys.\ {\bf 85}, 197 (2013).
%
%
%
%
%
\bibitem{Kaiser97} 
N.\ Kaiser, R.\ Brockmann, and W.\ Weise,
Nucl.\ Phys.\ A {\bf 625}, 758 (1997).
%
\bibitem{Kaiser98}
N.\ Kaiser, S.\ Gerstend\"{o}rfer, and W.\ Weise,
Nucl.\ Phys.\ A {\bf 637}, 395 (1998).
%
\bibitem{Kaiser99}
N.\ Kaiser, 
Phys.\ Rev.\ C {\bf 61}, 014003 (1999).
%
\bibitem{Kaiser00}
N.\ Kaiser, 
Phys.\ Rev.\ C {\bf 62}, 024001 (2000).
%
\bibitem{Kaiser01}
N.\ Kaiser, 
Phys.\ Rev.\ C {\bf 63}, 044010 (2001).
%
\bibitem{Kaiser011}
N. Kaiser, 
Phys.\ Rev.\ C {\bf 64}, 057001 (2001).
%
\bibitem{Kaiser02}
N.\ Kaiser, 
Phys.\ Rev.\ C {\bf 65}, 017001 (2002).
%
\bibitem{Epelbaum9800}
E.\ Epelbaum, W.\ Gl\"{o}ckle, and U.-G.\ Mei\ss{}ner, 
Nucl.\ Phys.\ A {\bf 637}, 107 (1998); A {\bf 671}, 295 (2000).
%
\bibitem{Entem02}
D.R.\ Entem and R.\ Machleidt,
Phys.\ Rev.\ C {\bf 66}, 014002 (2002).
%
\bibitem{Krebs07}
H.\ Krebs, E.\ Epelbaum, and Ulf.-G.\ Mei\ss{}ner,
Eur.\ Phys.\ J.\ A {\bf 32}, 127 (2007).
%
\bibitem{Entem03}
D.R.\ Entem and R.\ Machleidt,
Phys.\ Rev.\ C {\bf 68}, 041001(R) (2003).
%
\bibitem{Epelbaum05} 
E.\ Epelbaum, W.\ Gl\"{o}ckle, and U.-G.\ Mei\ss{}ner,
Nucl.\ Phys.\ A {\bf 747}, 362 (2005).
%
\bibitem{Siemens16}
D.\ Siemens, V.\ Bernard, E.\ Epelbaum, A.\ Gasparyan, H.\ Krebs, Ulf-G.\ Mei\ss{}ner,
 arXiv:1602.02640[nucl-th] (2016).
%
\bibitem{Yao16}
De-Liang Yao,  D.\ Siemens, V.\ Bernard, E.\ Epelbaum, A.M.\ Gasparyan, J.\ Gegelia, H.\ Krebs, Ulf-G. Mei\ss{}ner,
arXiv:1603.03638[nucl-th] (2016).
%
\bibitem{Entem15}
D.R.\ Entem, N.\ Kaiser, R.\ Machleidt, and Y.\ Nosyk,
Phys.\ Rev.\ C {\bf 91}, 014002 (2015). 
%
\bibitem{Epelbaum15}
E.\ Epelbaum, H.\ Krebs, U.-G.\ Mei\ss{}ner,
Phys.\ Rev.\ Lett. {\bf 115}, 122301 (2015).
%
\bibitem{Entem2015}
D.R.\ Entem, N.\ Kaiser, R.\ Machleidt, and Y.\ Nosyk,
Phys.\ Rev.\ C {\bf 92}, 064001 (2015).
%
\bibitem{Stoks93}
V.G.J.\ Stoks, R.A.M.\ Klomp, M.C.M.\ Rentmeester, and J.J.\ de Swart,
Phys.\ Rev.\ C {\bf 48}, 792 (1993).
%
\bibitem{Stoks94}
V.G.J.\ Stoks, R.A.M.\ Klomp, C.P.F.\ Terheggen, and J.J.\ de Swart,
Phys.\ Rev.\ C {\bf 49}, 2950 (1994).
%
\bibitem{SM99}
R.A.\ Arndt, I.I.\ Strakovsky, and R.L.\ Workman,
SAID, Scattering Analysis Interactive Dial-in computer
facility, George Washington University (formerly Virginia
Polytechnic Institute), solution SM99 (Summer
1999).
%
\bibitem{Navarro1314}
R.\ Navarro P\'erez, J.E.\ Amaro, and E.\ Ruiz Arriola,
Phys.\ Rev.\ C {\bf 88}, 064002 (2013);
Phys.\ Rev.\ C {\bf 89},  024004 (2014);
Phys.\ Rev.\ C {\bf 89}, 064006 (2014).
%
\bibitem{Ekstrom15}
A.\ Ekstr\"{o}m, G.R.\ Jansen, K.A.\ Wendt, G.\ Hagen, T.\ Papenbrock, B.D.\ Carlsson, C.\ Forss\'{e}n, M.\ Hjorth-Jensen, P.\ Navr\'{a}til, and W.\ Nazarewicz,
Phys. Rev. C 91, 051301(R) (2015).
%
\bibitem{Wiringa95}
R.\ B.\ Wiringa, V.\ G.\ J.\ Stoks, and R.\ Schiavilla,
Phys.\ Rev.\ C {\bf 51}, 38 (1995).
%
\bibitem{Carlson83}
J.\ Carlson, V.R.\ Pandharipande, and R.B.\ Wiringa,
Nucl. Phys. A {\bf 401}, 59 (1983).
%
\bibitem{Pudliner96}
B.S.\ Pudliner, A.\ Smerzi, J.\ Carlson, V.R.\ Pandharipande, Steven C.\ Pieper, and D.G.\ Ravenhall,
Phys.\ Rev.\ Lett. {\bf 76}, 2416 (1996).
%
\bibitem{Pieper08}
S.C.\ Pieper,
AIP Conf. Proc. {\bf 1011}, 143 (2008).
%
\bibitem{Pieper01}
S.C.\ Pieper, V.R.\ Pandharipande, R.B.\ Wiringa, and J.\ Carlson,
Phys.\ Rev.\ C {\bf 64}, 014001 (2001).
%
\bibitem{Gezerlis13}
A.\ Gezerlis {\it et al.},
Phys.\ Rev.\ Lett. {\bf 111} (3), 032501 (2013).
%
\bibitem{Gezerlis14}
A.\ Gezerlis, I.\ Tews, E.\ Epelbaum, M.\ Freunek, S.\ Gandolfi, K.\ Hebeler, A.\ Nogga, and A.\ Schwenk,
Phys.\ Rev.\ C {\bf 90}, 054323 (2014).
%
\bibitem{Lynn14}
J.E.\ Lynn, J.\ Carlson, E.\ Epelbaum, S.\ Gandolfi, A.\ Gezerlis, and A.\ Schwenk,
Phys.\ Rev.\ Lett.\ {\bf 113}, 192501 (2014).
%
\bibitem{Lynn16}
J.E.\ Lynn, I.\ Tews, J.\ Carlson, S.\ Gandolfi, A.\ Gezerlis, K.E.\ Schmidt, and A.\ Schwenk,
Phys.\ Rev.\ Lett.\ {\bf 116}, 062501 (2016).
%
\bibitem{Piarulli15}
M.\ Piarulli, L.\ Girlanda, R.\ Schiavilla, R.\ Navarro P\'{e}rez, J.E.\ Amaro, and E.\ Ruiz Arriola,
Phys.\ Rev.\ C {\bf 91}, 024003 (2015).
%
\bibitem{Long11}
B.\ Long and V.\ Lensky,
Phys.\ Rev.\ C {\bf 83}, 045206 (2011).
%
\bibitem{POUNDerS} 
M.\ Kortelainen, T.\ Lesinski, J.\ Mor\'{e}, W.\ Nazarewicz, J.\ Sarich, N.\ Schunck, M.V.\ Stoitsov, and S.\ Wild,
Phys.\ Rev.\ C {\bf 82}, 024313 (2010).
%
\bibitem{Gross08}
F.L.\ Gross and A.\ Stadler,
Phys.\ Rev.\ C {\bf 78}, 014005 (2008).
%
\bibitem{Bergervoet88}
J.\ R.\ Bergervoet, P.\ C.\ van Campen, W.\ A.\ van der Sanden, and J.\ J.\ de Swart, 
Phys. Rev. C {\bf 38} (1988) 15.
%
\bibitem{Sanden83}
W.\ A.\ van der Sanden, A.\ H.\ Emmen, and J.\ J.\ de Swart, 
Report No. THEF-NYM-83.11, Nijmegen (1983), unpublished; quoted in~\cite{Bergervoet88}.
%
\bibitem{Chen08}
Q.\ Chen {\it et al}., 
Phys.\ Rev. {\bf C} 77 (2008) 054002.
%
\bibitem{Miller90}
G.\ A.\ Miller, M.\ K.\ Nefkens, and I.\ Slaus, 
Phys.\ Rep. {\bf 194} (1990) 1.
%
\bibitem{Machleidt01}
R. Machleidt, 
Phys.\ Rev.\ C {\bf 63} (2001) 024001.
%
\bibitem{Ericson83}
T.\ E.\ O.\ Ericson and M.\ Rosa-Clot, 
Nucl.\ Phys.\ A {\bf 405}, 497 (1983).
%
\bibitem{Rodning90}
N.\ L.\ Rodning and L.\ D.\ Knutson, 
Phys.\ Rev.\ C {\bf 41}, 898 (1990).
%
\bibitem{Huber98}
A.\ Huber, T.\ Udem, B.\ Gross, J.\ Reichert, M.\ Kourogi, K.\ Pachucki, M.\ Weitz and T.\ W.\ Hansch, 
Phys. Rev. Lett. {\bf 80}, 468 (1998).
%
\bibitem{Martorell95}
J.\ Martorell, D.\ W.\ L.\ Sprung and D.\ C.\ Zheng, 
Phys.\ Rev.\ C {\bf 51}, 1127 (1995).
%
\bibitem{Piarulli13}
M.\ Piarulli, L.\ Girlanda, L.E.\ Marcucci, S.\ Pastore, R.\ Schiavilla, and M.\ Viviani,
Phys.\ Rev.\ C {\bf 87}, 014006 (2013).
%
\bibitem{Kie94}
A.\ Kievsky, M.\ Viviani, and S.\ Rosati, 
Nucl.\ Phys.\ A {\bf 577}, 511 (1994).
%
\bibitem{Kie97}
A.\ Kievsky, L.E.\ Marcucci, S.\ Rosati, and M.\ Viviani, 
Few-Body Syst.\ {\bf 22}, 1 (1997).
%
\bibitem{Viv05}
M.\ Viviani, A.\ Kievsky, and S.\ Rosati, 
Phys.\ Rev.\ C {\bf 71}, 024006 (2005).
%
\bibitem{Kie08}
A.\ Kievsky, S.\ Rosati, M.\ Viviani, L.E.\ Marcucci, and L.\ Girlanda,
           J.\ Phys.\ G: Nucl.\ Part.\ Phys.\ {\bf 35}, 063101 (2008).
%
\bibitem{Metropolis53}
N.\ Metropolis, A.\ W.\ Rosenbluth, M.\ N.\ Rosenbluth, A.\ H.\ Teller, and E.\ Teller,
J.\ Chem.\ Phys. {\bf 21}, 1087 (1953).
%
\bibitem{Wiringa91}
R.B.\ Wiringa, 
Phys.\ Rev.\ C {\bf 43}, 1585 (1991).
%
\bibitem{Carlson87}
J.\ Carlson,
Phys.\ Rev.\ C {\bf 36}, 2026 (1987).
%
\bibitem{Carlson88}
J.\ Carlson, 
Phys. Rev. C {\bf 38}, 1879 (1988).
%
\bibitem{NLopt}
http://ab-initio.mit.edu/wiki/index.php/NLopt
%
\bibitem{Lynn12}
J.E.\ Lynn and K.E.\ Schmidt,
Phys.\ Rev.\ C {\bf 86}, 014324 (2012).
%
\bibitem{Pudliner97}
B.S.\ Pudliner, V.R.\ Pandharipande, J.\ Carlson, Steven C.\ Pieper, and R.B.\ Wiringa,
Phys.\ Rev.\ C {\bf 56}, 1720 (1997).
%
\bibitem{Wiringa00}
R.B.\ Wiringa, Steven C.\ Pieper, J.\ Carlson, and V.R.\ Pandharipande,
Phys.\ Rev.\ C {\bf 62}, 014001 (2000).
%
%
\end{thebibliography}
\end{document}